%% file: main.tex
\documentclass[conference,letterpaper]{IEEEtran}

\usepackage{float}

\usepackage{soul}
\usepackage{epsfig}
\usepackage{graphicx}
\usepackage{amsmath}
\usepackage{amssymb}
\usepackage{algorithm}
\usepackage{algpseudocode}%
\usepackage{url}
\usepackage[tight,footnotesize]{subfigure}
\usepackage{multirow}
\usepackage{balance}
\usepackage{array}
\usepackage{color}
\usepackage{bm}
\usepackage[normalem]{ulem}
\usepackage{authblk}

\usepackage{graphicx,subfigure}

\author[1]{Duc-Ly Vu}
\author[2]{Trong-Kha Nguyen}
\author[3]{Tam V. Nguyen}
\author[4]{Tu N. Nguyen}
\author[1]{Fabio Massacci}
\author[3]{Phu H. Phung}
\affil[1]{Department of Information Engineering and Computer Science, University of Trento, Italy \authorcr Email: {\tt \{ducly.vu, fabio.massacci\}@unitn.it}\vspace{1ex}}
\affil[2]{Hongik University, South Korea \authorcr Email: {\tt \{nguyentrongkha92\}@gmail.com}\vspace{1ex}}
\affil[3]{Department of Computer Science, University of Dayton, U.S.A. \authorcr Email: {\tt \{tamnguyen, phu\}@udayton.edu}\vspace{1ex}}

\affil[4]{Department of Computer Science, Purdue University Fort Wayne, U.S.A. \authorcr Email: {\tt nguyent@pfw.edu }}

\begin{document}

\title{A Convolutional Transformation Network for Malware Classification}


\maketitle

\thispagestyle{plain}
\pagestyle{plain}

\begin{abstract}


Modern malware evolves various detection avoidance techniques to bypass the state-of-the-art detection methods. An emerging trend to deal with this issue is the combination of image transformation and machine learning techniques to classify and detect malware. However, existing works in this field only perform simple image transformation methods that limit the accuracy of the detection. In this paper, we introduce a novel approach to classify malware by using a deep network on images transformed from binary samples. In particular, we first develop a novel hybrid image transformation method to convert binaries into color images that convey the binary semantics. The images are trained by a deep convolutional neural network that later classifies the test inputs into benign or malicious categories. Through the extensive experiments, our proposed method surpasses all baselines and achieves 99.14\% in terms of accuracy on the testing set.

\end{abstract}




\section{Introduction}
\label{sec:intro}
\input{parts/intro.tex}


\section{Related Work}
\label{sec:related}
\input{parts/relatedwork.tex}

\section{System design}
\label{sec:method}
\input{parts/methodology}

\section{Experiments}
\label{sec:experiments}

\label{sec:sysmodel}
\input{parts/sysmodel}

\section{Conclusions}
\label{sec:conc}

In this paper, we present a Convolutional Transformation Network for malware classification based on the combination of deep learning and the conversion of binary files into color images. In other words, we cast the malware classification problem into the image classification task. We improve the accuracy rate by enhancing the image color coding. The results of malware classification show that our method achieves over 99.14\% regarding accuracy surpassing all the baselines.

In the future, we would like to extend our work to the malware segmentation problem to detect the specific malicious segments inside malware programs. Also, we also aim to investigate our work to polymorphism malware classification.


\section*{Acknowledgements}

The project leading to this paper has received funding from the
European Union’s Horizon 2020 research and innovation programme under grant agreement No 675320 (NeCS: European Network for Cyber Security).







\balance

\bibliographystyle{IEEEtran}
\bibliography{references}
\balance
\end{document}

%% file: parts/intro.tex
The number of new malware and variants on the Internet has been continuously increasing. According to a technical report from Kaspersky Lab~\cite{KasperskyLab:2016}, 323,000 new malware files are detected daily, and there are one billion unique malware files in their cloud database. This significant growth is due to the existing of many automatic malware creation toolkits such as Zeus and SpyEye~\cite{MalDet:Surveys:2017}. These kits commonly employ different evading techniques such as encryption, obfuscation, packing to create new malware from existing malware samples to bypass the detection of anti-malware scanners~\cite{MalDet:Surveys:2017}.

In recent years, machine learning and deep learning techniques have been adopted to the malware classification domain by researchers and anti-malware vendors to capture the malware evading techniques~\cite{MalDet:Surveys:2017,evading:2017} (the next section discusses the detailed related work). Machine learning algorithms in malware classification are based on a set of features extracted from file samples by static and/or dynamic analysis techniques. The analyses require either the disassembly code or code execution, and the accuracy of these models is, therefore, dependent on the analysis tools and selected features from the analyses. Also, the analysis and feature selecting process sometimes need security experts to revise and disambiguate intermediate results, thus cannot be fully automated~\cite{KasperskyLab:2016,VisualizationSystems:Survey:2015}. 

An alternative approach to overcoming the limitations mentioned above was the adoption of image processing and classification techniques to the domain of malware. The principal tenet of this approach is that it does not require disassembly code or code execution since binary files are converted and mapped to images so that it can be resilient to the known anti-analysis techniques~\cite{VisualizationSystems:Survey:2015,nataraj2011malware}.

Although image-based is a new approach for malware classification that can avoid anti-analysis techniques, the existing works in this domain e.g.,~\cite{nataraj2011malware,nataraj2016SPAM,Makandar2017ICDMAI,liu2016malware,han2015malware,shaid2014malware} use simple mapping algorithms to transform malware binaries to images. Thus the semantics of the malware may be disregarded. Our observation is that the more information given to classifiers, the more accuracy rate can be archived. Motivated by this, our work proposes a novel hybrid image transformation method for binaries for malware classification. Our new technique tackles the semantic issue by adding and highlighting essential sequences visually using the entropy technique. A binary file is transformed into a color image where its channels encode semantic information. By visualizing continuous sequences of same semantic entropy values, the image can highlight suspicious sections in a binary for further analysis such as packed or encrypted sections or small cryptographic artifacts like decryption keys or passwords which meant to be hidden from human view. Our approach starts with a simple color scheme where bytes are classified into a small number of categories to get an overview of the structure of a file. This approach allows us to select the best color scheme to represent the semantic meaning of the binary data, Then, we take the byte stream and split them into 32-bit blocks and calculate the Shannon entropy on them. Using the second scheme, we can locate encrypted or packed sections. Taken from 256 different byte values, we compress them down into a few common character classes and calculate byte entropy over a sliding window of these selected bytes. Each of these color schemes has its own advantages in representing malware behavior. The character class scheme covers the most common padding bytes, nicely highlights strings in malware while the entropy scheme will locate encrypted and compressed sections. 

To automatically recognize malware variants, their shared patterns should be identified and learned. If malware authors make a small change in the original binary, its image retains the global structure~\cite{MalDet:Surveys:2017}. Thus, image representations of different binaries from the same malware family appear to be similar. Based on these observations, we have developed a deep neural network to learn the global patterns shared among malware. 

Our proposed method consists of the following steps in a supervised training phase. First, we divide a given binary executable (benign or malicious) into blocks of sequence and calculate the Shanon entropy value for each block. Next, these blocks are analyzed conditionally and assigned to the corresponding color. The transformed images are fed into convolutional/pooling/fully connected layers. Finally, the network outputs the predicted label. In summary, the contributions of our work are as follows.

\begin{itemize}
\item We propose a novel image transformation method to convert binary executables into color images that convey the semantics of the binary data.
\item We develop a Convolutional Transformation Network, or CTN in short, for classifying malware based on transformed images. 
\item We report the experimental results of our proposed network on a large dataset of malicious and benign binary samples. Through extensive experiments, our proposed method achieves 99.14\% of accuracy, a much greater rate compared with similar work on the same test set. 
\end{itemize}

The composition of this paper is as follows: Section~\ref{sec:related} describes related work on malware analysis, detection, and classification methods. In Section~\ref{sec:method}, a malware analysis using color images is proposed, and Section~\ref{sec:experiments} illustrates the experimental results. Finally, the conclusions of this paper are presented, and future work is discussed in Section~\ref{sec:conc}.  

\begin{figure*}[h!]
	\centering
	\includegraphics[width = .875\linewidth]{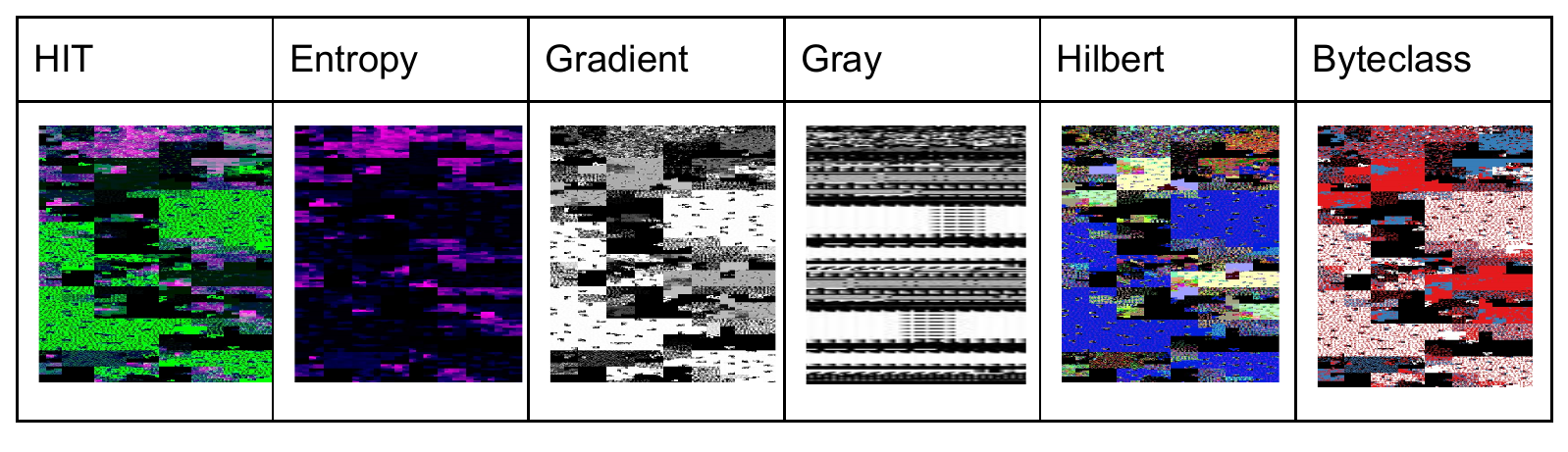}
	\caption{Visualization of a sample using different transformation schemes.}
	\label{fig:system_model}
\end{figure*}

%% file: parts/relatedwork.tex
In this section, we first review the progress of image classification. Then, we summarize the malware classification and the integration of the image transformation into the malware classification.

\subsection{Image classification}

Image classification is a fundamental problem in computer vision. In the early stage, Haralick \textit{et al.} \cite{haralick1973textural} describes some easily computable textural features based on gray-tone spatial dependencies and illustrates their application in category-identification tasks. Later, LeCun \textit{et al.} \cite{lecun1998gradient} proposed Convolutional neural networks (ConvNets or CNNs) by stacking several convolutional operators into a network. CNNs can create a hierarchy of progressively more abstract features and show a good performance on hand-writing digit classification. However, the limitation of hardware resource has restricted CNNs from further investigation. Therefore, there exist many works using hand-crafted features for image classification. In the global scale, GIST~\cite{oliva2001modeling} is computed over the entire image as a global image descriptor for scene classification. In the local scale, Lowe~\cite{SIFT} introduced SIFT feature extracted from interest points. Similarly, Dalal \textit{et al.}~\cite{HOG} proposed histogram of gradients (HOG) which can be used for both image classification and object detection in a sliding window manner. 

Recently, along with the development of GPUs, CNNs were resurrected in deep learning for image classification. Krizhevsky \textit{et al.}~\cite{krizhevsky2012imagenet} trained a large, deep convolutional neural network to classify the 1.2 million high-resolution images in the ImageNet LSVRC-2010 contest into the 1000 different classes. The network has many layers, namely, convolution layers and max-pooling layers, and fully-connected layers with a final 1000-class softmax. Furthermore, there are many extensions to deeper networks for higher performance in terms of classification rate~\cite{VGG,Googlenet,ResNet}. 

\subsection{Malware Classification }
Malware classification is a method to detect whether a
given software program is malicious. Conventional techniques
in the anti-malware industry used signature-based algorithms,
however, these techniques cannot detect new malware or modified
malware using evading techniques such as encryption,
packing, polymorphism, obfuscation, and meta-morphism~\cite{MalDet:Surveys:2017,evading:2017}. To detect these new type of malware, in recent years, there
have been various efforts to adopt advanced machine learning
techniques in the domain of malware classification. In this
subsection, we summary latest efforts in this area. We also
highlight several works closely related our work that adopted
image classification techniques to detect malware.

\subsubsection{Feature-based Approach} These classification techniques
first extract various features from file samples and
use these features to train the classifier using machine learning
methods. Feature extraction can be performed by static
analysis, dynamic analysis, or a hybrid combination of the
two. Static analysis techniques performed a string search on
the program to collect some features. Many efforts used static analysis to construct a feature vector for
classification such as \cite{kolter2006learning,schultz2001data,tian2009automated,islam2013classification,dahl2013large,saxe2015deep}. The limitations of static analysis are that static-based features
suffer from binary obfuscation and are limited in representing
true behavior of malware. Moreover, these techniques are
platform-specific and only applicable to Windows PE. Various
other works used dynamic techniques to extract features from
operations on system resources~\cite{rieck2008learning,huang2016mtnet}, call sequences \cite{SystemCallSequences:2016},
control flow or function call graph \cite{Autoencoder-based:2017}. According to recent
findings, e.g., \cite{evading:2017,FeatureSqueezing:2018,EvadingClassifiers:2016} features-based malware classification
methods are still facing evading techniques implemented
in modern malware.

\subsubsection{Image-based Approach} There have been several efforts
that adapt the techniques in image processing to malware
classification. In \cite{MalDet:Surveys:2017,VisualizationSystems:Survey:2015}, various visualization techniques, 
including image processing for malware analysis, have been
surveyed. Lately, more efforts are adopting the image-based
approach for malware classification, e.g., \cite{Makandar2017ICDMAI,liu2016malware,han2015malware,shaid2014malware}. A
common feature of these efforts is that they transform binary malware samples into different image forms, then image
classification techniques are used to classify the malware based
on the image representation of the malware. We highlight a
few works closely related to ours and categorize them based
on the image transformation methods and the representation
of the malware in images.
In \cite{nataraj2016SPAM,nataraj2011comparative}, the binary samples are mapped to a vector of
8-bit unsigned integers of byte values, which is reshaped and converted into a gray-scale image in the range [0,255]. The
malware classification is calculated using computed texture
features from the images using K-nearest neighbors.
Han et al. \cite{han2013malware,han2014malware} proposed several transformation methods
to convert the opcode sequences extracted from malware
samples into image matrices represented in RGB-colored
pixels. In a later work of the same authors \cite{han2015malware}, binaries are
converted into bitmap images, which are then converted to
entropy graphs to calculate the similarities. This approach
however, only works for Windows PE file because it needs PE
header information to decide sections to be converted. Besides,
this method cannot deal with packed samples.
Liu et al. \cite{liu2016malware} proposed a method to transform disassembly
files to gray-scale images, which are later compressed and
mapped into feature vectors for classification using K-means
and diversity selection. In \cite{shaid2014malware}, malware is executed on a
virtual machine to capture user-mode API calls, which are then
sorted assigned a color based on their maliciousness. The most
malicious APIs such as DeleteFileA are assigned hot colors
(stars from red (1,0,0) in RGB) while the most benign APIs
are mapped to cold colors (the coldest color is blue (0,0,1)
in RGB). The API color points at the time they appear in
log trace are rendered into an image that is used to classify
the maliciousness. Kancherla \textit{et al.} \cite{kancherla2013image} plots raw bytes values
into 2-D dimensional images and extract intensity, Wavelet,
and Gabor-based features. These features are then fed into the
SVM classifier to do malware classification.
Our work stands apart from the literature by introducing a novel transformation method to convert binaries to images with multiple layers. Thanks to this approach, more features
of binaries are carried in the images, resulting in a higher accuracy rate of classification compared with existing works.

%% file: parts/methodology.tex
In this section, we introduce our proposed Convolutional Transformation Network (CTN) for malware classification. Our network consists of two major components, namely, input transformation and convolutional network-based classification.

\begin{figure*}[!thb]
    \centering
    \includegraphics[width = \linewidth]{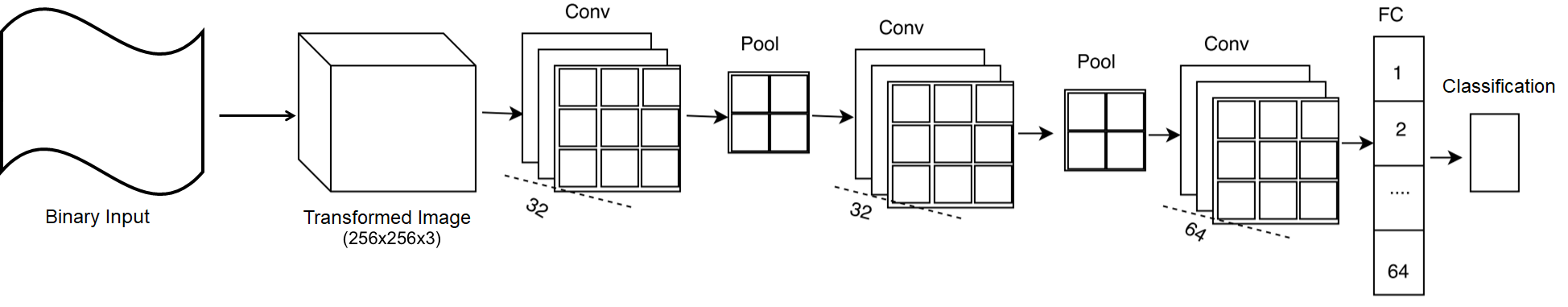}
    \caption{Pineline of our Convolutional Transformation Network for malware classification. }
    \label{fig:system_model}
\end{figure*}
\subsection{Motivation}
Malware uses obfuscation technique to bypass Antivirus and hide its malicious activities. To better capture their behavior, we not only use static features like \textit{strings}, \textit{imports} but also need to understand the encoding techniques. The output of the malware detection system can give a better intuition of malware to users rather than giving a single decision. In other words, it is beneficial to point out suspicious section in a program, and security analysts can perform further analysis on them. To achieve these goals, we first analyze the raw byte contents of programs and split it into blocks of byte sequences. We then calculate entropy for each block and transform it into color images. The color images have been proven to be more effective than the gray scale counterpart by the work~\cite{ColorSIFT}. 

\subsection{Input Transformation}

In this section, we explore multiple ways to encode and transform the binary input into images. First, we start with a simple technique called the \textit{Byte class}

\subsubsection{Byte class}
This scheme only includes information about  \textit{strings}. Specifically, a character belongs to one of the four categories: the lowest byte value (0), the highest, lowest byte value, the printable strings, and non-printable strings. For example Tab (09), newline (0a) and carriage return (0d) are considered to be texts. This method can be used to get an overview of the file structure.

\subsubsection{Gradient based}
Gradient color scheme is similar to the byteclass method except that they vary colors with byte ordinal values, which is from 0 to 255.

This scheme can reveal structural details that do not appear in the byteclass scheme.

\subsubsection{Hilbert}
The detail colors scheme assigns a color to each different byte value. It tries to maximize the difference between colors, while at the same time keeping colors for bytes that are close in value as similar as possible. 
To balance these two conflicting constraints, we again resort to the Hilbert curve. In this paper, we project the 1-dimensional sequence of byte values into a 3-dimensional Hilbert curve traversal of RGB color cube.  

\subsubsection{Entropy}
Entropy  is used to detect packed or encrypted malware~\cite{lyda2007using}. 
In this method a file is read from the beginning, and divided into blocks of byte sequences. Using the Shannon Entropy~\cite{shannon1951prediction}: 
\begin{equation}
H(X)= -\sum_{i=0}^{N-1}p_ilog_bp_i 
\end{equation}
where $X$ is a random variable of 256 symbol map of $N$; ${i = 0,\dots, 255}$; block size  $b$ is the block size. The probability mass function, $p_i$ is the probability of byte value i within a given byte block. Entropy values are then rescaled to between 0 and 1. We assign entropy values to printable and encoded strings, and transform them into color RGB images.

\begin{figure}
\centering
\includegraphics[scale=0.3]{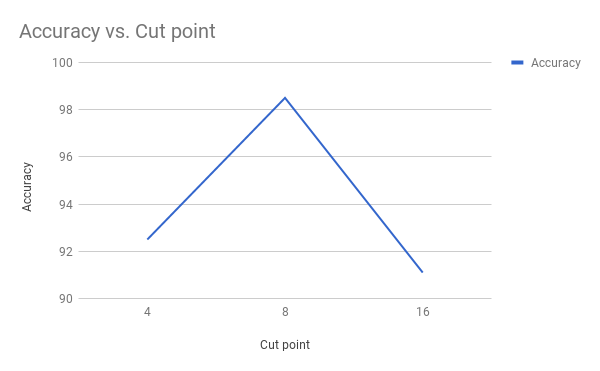}
\caption{Detail selection}
\label{fig:selectiondetail}
\end{figure}

\begin{table*}[!ht]
\centering
\normalsize
\caption{Performance comparison. The best performance of each category is highlighted with boldface font.}
\label{compare_table}
\begin{tabular}{|l|c|c|c|c|}
\hline
Scheme     & Training (GIST)~~~~~ & Validation (GIST) & Training (CNN)~~~~~ & Validation (CNN)  \\ \hline
Class &   90.78   & 81.81 & 94.50 & 94.40\\ \hline
Gray       & \textbf{93.57}  &  \textbf{94.27} & 97.62 & 95.31  \\ \hline
Hilbert    & 88.88  & 82.32 & 93.42 & 96.61\\ \hline
Gradient~~~~~~~~~~    &  88.41   & 82.91 & 97.57 &93.23\\ \hline
Entropy    & 88.76    & 77.77 & 97.42 & 93.88\\ 
\hline
HIT    & 91.15      & 84.34 & \textbf{98.82}& \textbf{99.14}\\ \hline

\end{tabular}
\end{table*}
\subsection{Hybrid Image Transformation (HIT)}

We extend the entropy color scheme in the previous section to capture more semantic information about PE files. By encoding byte values in a wide range, HIT can not only capture obfuscation information but also be able to represent semantic information in the file headers like imported functions or libraries.
We encode the semantic information into the green channel of RGB image. To select the best number of partitions in the 256 symbols, we base on the observation that information in the PE headers exists in a visible form of strings and numbers. We start by splitting the symbol range into 4 smaller ranges of lowest, highest, printable and other bytes, and keep splitting the range by a binary value $2^n$ where $n$ is the index of partition until getting the best performance. We reserve the red and blue as same as entropy method and put more light to green channel on standard characters. Table~\ref{tab:bytecolor} illustrates a typical example of byte splitting and encoding. The motivation behind our color scheme is that a detailed image can improve classification performance~\cite{van2010evaluating} and by using more ranges of byte values metadata information such as PE headers, in an executable can be visually identified. As almost malware samples are packed to hinder the analysis, HIT can be applied to detect not only obfuscation patterns and malicious indicators in executables. A regular binary has lower entropy than a packed binary since its follow a software coding standard and contains printable characters. 
Packed malware has higher entropy than benign since obfuscated or packed makes bytes randomly. 

We store entropy information into the red and blue channels while put string information into the green channel as it is most sensitive to human vision~\cite{fortner2012number} and take highest coefficient value in image grayscale-color conversion. In particular, our HIT method defines entropy value for red and blue and fixed green value by a binary value $2^n$, where $n$ is the index of partition.  Thus, HIT can output a lower entropy value that makes red and blue lower, and a pixel tends to green. By this way, regular files have more green pixel than malicious files, which contain higher entropy due to the higher red/blue values.


Designing HIT, however, requires selecting the number of partitions. If we divide the color range into many partitions, the output pixels in the image are going to be random since the patterns were removed. Furthermore, when the number of partitions increases, the classifiers learn patterns from an image because it contains random pixels or the training process are easier to be overfitted. We do a heuristic search for the best cut.  

\begin{table}[!thb]
\renewcommand{\arraystretch}{1.3}
\caption{Mapping table for 8 ranges }
\label{tab:bytecolor}
\centering
\begin{tabular}{|l|c|}
\hline
\textbf{Byte value} & \textbf{Bytes}\\
\hline
0 & RGB(r, 0, b) \\ \hline
255 & RGB(r, 255, b) \\ \hline
[a-w] & RGB(r, 126, b) \\ \hline
[A-W] & RGB(r, 64, b) \\ \hline
[0-9] & RGB(r, 32, b) \\ \hline
special character & RGB(r, 16, b) \\ \hline
\end{tabular}
\end{table}

\subsection{Network Layers}
There exist standard convolutional/pooling layers that widely used in image classification such as AlexNet~\cite{krizhevsky2012imagenet}. In our work, we design a simple neural network which can be used to extract features from images. In particular, our CTN model contains three convolutional layers, two pooling layers, and one fully connected layer, respectively, to learn to classify transformed images automatically. Note that the input of our network is the transformed images, as discussed earlier. Fig.~\ref{fig:system_model} represents network operators used in our model. 


%% file: parts/sysmodel.tex
\subsection{Dataset}
For the evaluation, we collected malware samples from
Virusshare~\cite{virusshare} and Windows executable software as benign
files. We utilize the Microsoft Software Removal Tool \cite{msantivirus} to
label the samples. Our dataset contains 525 malicious
and 525 benign selected samples. We further partition the
dataset into two parts: the training set (80\%), the validation
set (20\%).
\subsection{Experimental Results}
We first conduct the parameter selection. As shown in
Fig.~\ref{fig:selectiondetail}, our proposed model achieves the best performance
at the cut point 8. In case we increase the cut point to 16, the performance of the model is decreasing as more random
pixels cannot be learned well. Also, it slows down
and overfits classifiers. Therefore, we adopt the cut point 8 for
the rest of our experiments.

Table~\ref{compare_table} demonstrates the performance of different image transformation methods
and different image features. CNN
generally performs better than GIST. Our HIT performs
the best on both training set and test set. That means the
HIT can be able to generalize well on unseen samples. The
performance on the training set is generally better than the one
on the validation. There is one exception that the entropy
transformation performs better with new unseen data on CNN.
Regarding the GIST feature, the gray image transformation
performs the best with 94.27\%. While using CNN features,
HIT reaches the top performance with \textbf{99.14\%} accuracy.
The results indicates that using color images with CNN
architecture is better than gray scale images.
\begin{figure}
\centering     
\subfigure[Benign average]{\label{fig:a}\includegraphics[width=40mm]{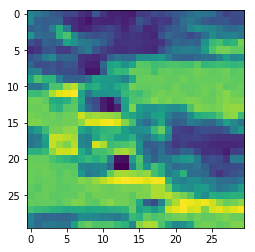}}
\subfigure[Malicious average]{\label{fig:b}\includegraphics[width=40mm]{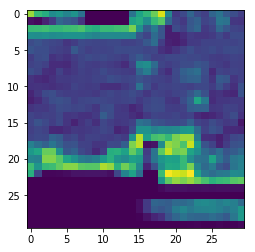}}
\caption{Mean output of benign (left) and malicious (right) samples.}
\label{fig:meanviz}
\end{figure}

Fig.~\ref{fig:meanviz} visualizes the output average taken from the last
network in our model. We can observe that the mean outputs
of benign and malicious samples are different. Fig. \ref{fig:a} shows that benign samples in average have almost locations in
the green regions while malicious samples have more sections
in the dark blue regions, as shown in Fig.~\ref{fig:b}. It obviously
indicates that malware is evolved with obfuscation techniques
to hide visible indicators.